# High-Gain Voltage-Multiplier Coupled Quadratic Boost Converter: A New Design for Small Scale PV Integration

Safa Mohammed Sali, Hoach The Nguyen, *Senior Member*, *IEEE*, Ameena Saad Al-Sumaiti, *Senior Member*, *IEEE*.

*Abstract*—This paper introduces a single-switch high-gain voltage-multiplier coupled quadratic boost converter (HGVM-QBC), developed from the conventional quadratic boost converter (QBC). The proposed topology is designed to achieve higher voltage gain, lower semiconductor voltage stress, and continuous current operation, making it particularly suitable for small-scale photovoltaic (PV) systems. By incorporating a voltage multiplier cell into the QBC, the converter significantly improves voltage boosting capability while mitigating stress on switching devices. In this configuration, the output voltage is obtained by combining the voltages across multiple output capacitors, thereby enhancing the overall voltage level. A detailed comparative study with recently reported converter topologies demonstrates the superior gain and reduced device stress offered by the HGVM-QBC. The design is validated through MATLAB/Simulink simulations, which confirm improved performance in terms of gain and voltage stress. Furthermore, an experimental prototype achieves an output of 151 Vdc from a 12 Vdc input at a 55% duty cycle, corresponding to a gain of 12.59. These results establish the HGVM-QBC as an efficient and reliable solution for PV applications that demand high voltage output from low input sources.

## I. INTRODUCTION

Recently, in response to climate change, the need to reduce greenhouse gas emissions, and the decreasing reliance on fossil fuels, there has been a significant global shift toward renewable energy sources (RES) [1]. Among RES, photovoltaic (PV) systems have gained wide adoption due to their environmentally friendly nature and declining cost [2]. PV systems are typically classified as either stand-alone [3] or grid-connected [4]. Stand-alone systems may or may not include energy storage and are generally used in small-scale applications. Since energy storage represents the major cost component, investment considerations drive this classification. Grid-connected PV systems avoid costly storage, feeding power directly into the grid, and are categorized into small, medium, and utility-scale applications. In contrast, stand-alone systems rely on bulky, expensive, and short-lifespan batteries, which make grid-connected systems more attractive for large-scale deployment. Nevertheless, small-scale PV systems, typically below 10 kW, remain popular. However, the voltage output from PV cells is inherently low and unsuitable for direct grid connection without additional voltage boosting [5].

To address the issue of low DC voltage, DC–DC step-up converters are widely used to increase PV output voltages, supplementing the series connection of PV cells [6]. These converters act as intermediaries between the PV system and the grid, elevating the low source voltage to higher DC levels [3], [7], as shown in Fig. 1. DC–DC converters are generally classified into isolated and non-isolated types.

Isolated converters use transformers to achieve gain through the turns ratio, but they suffer from leakage inductance, large size, and high cost [8]. Non-isolated converters, by contrast, provide higher efficiency due to lower losses, compact design, faster dynamic response, and reduced cost [9].

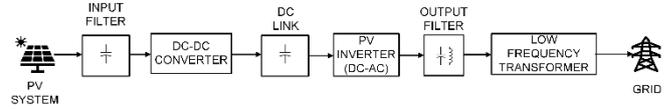

Fig. 1. Structure of grid-connected PV systems.

Common examples include boost, buck–boost, Cuk, Zeta, Luo [10], quadratic boost, and single-ended primary inductor converters (SEPIC). Despite their popularity, conventional converters face challenges such as high semiconductor voltage stress and limited voltage gain, reducing their efficiency in small-scale PV applications.

Improving voltage gain while simultaneously reducing semiconductor voltage stress is therefore a critical design goal in DC–DC converters [11]. Various strategies have been explored. For example, a bidirectional switched quasi-Z-source converter with common ground was proposed in [12], where repositioning the main switch simplifies the topology and allows use of low-voltage, low-resistance devices. In [13], the classical boost converter was modified with a charging circuit and an intermediate capacitor, reducing stress without altering the conversion ratio. Zero-voltage switching (ZVS) techniques have been applied to reduce switching losses [14], though this does not improve gain or reduce stress. Other approaches involve passive switched-capacitor (PSC) networks [15], or active switched-inductor (ASL) networks, where inductors are charged in parallel during the switch-on state and discharged in series during the switch-off state, forming an ASL step-up two-cell (ASL-SU2C) converter [16], [17]. Voltage stress in switched inductor boost converters has also been mitigated by adding auxiliary switches [18]. Advanced methods such as the double-duty triple-mode (DDTM) converter [19] or multi-switch architectures [20]achieve higher gains and lower stresses by exploiting multiple duty ratios and energy transfer loops. Coupled inductors with voltage multiplier cells [21] and hybrid topologies combining switched-capacitor and conventional converters [22] further enhance gain and efficiency. Recycling leakage inductance also contributes to stress reduction [23].

Among non-isolated converters, the quadratic boost converter (QBC) has distinct advantages. It achieves high gain at relatively low duty cycles, making it suitable for high-voltage, low-power applications [9]. Compared with Cuk, Zeta, and SEPIC converters, QBCs impose lower voltage stress on switches and diodes [24] . Unlike buck–boost and Cuk designs, they also produce a non-inverted output and operate in continuous current mode (CCM), which improves control and stability [25], [26].

Nonetheless, the conversion ratio of a conventional QBC is often insufficient for applications requiring very high voltage boosting. Several modifications have therefore been proposed. For example, introducing an additional switch while removing the coupled inductor improves gain and reduces stress [27]. Voltage multiplier cells (VMCs) integrated with the QBC further enhance both efficiency and gain, while reducing semiconductor stress[28], [29].

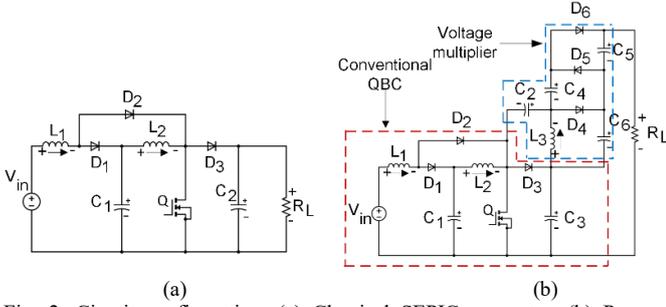

Fig. 2. Circuit configuration. (a) Classical SEPIC converter. (b) Proposed enhanced gain SEPIC converter (HGVM-QBC).

Simpler designs with fewer components have also been shown to lower inductor ripple and device stress [30]. Despite these efforts, most modified QBCs still suffer from limited gain and significant voltage stress.

To overcome these limitations, this paper proposes an advanced quadratic boost converter: the single-switch high-gain voltage-multiplier coupled quadratic boost converter (HGVM-QBC). Fig. 2 illustrates the topologies of both the conventional QBC and the proposed HGVM-QBC. In conventional QBCs, voltage stress equals the sum of the input and output voltages, limiting efficiency. By contrast, the proposed design introduces a voltage multiplier cell and non-coupled inductors, yielding the following advantages:
- High gain and output voltage at low duty cycles while operating in CCM, making it well-suited for small-scale PV systems.
- Reduced voltage stress on both switch and diodes, enabled by the single-switch architecture and VMC integration.
- Enhanced gain by increasing the numerator of the voltage gain equation through the VMC.
- Further reduced stresses at higher duty ratios due to the use of non-coupled inductors, improving overall efficiency and performance.

II. STEADY-STATE ANALYSIS OF THE HIGH GAIN AND LOW VOLTAGE STRESS CONVERTER

This section presents a novel DC–DC converter topology that achieves high voltage gain and reduced semiconductor stress using a single switch. The proposed design integrates a voltage multiplier (VM) network with the conventional quadratic boost converter (QBC), thereby enhancing the voltage conversion ratio while simultaneously lowering the stress on the switching devices.

In this topology, the output voltage is obtained by summing the voltages across multiple output capacitors, resulting in a significantly higher overall output level. The converter is configured to operate in continuous conduction mode (CCM) to ensure stable current flow and improved efficiency.

The steady-state operation of the proposed converter can be divided into four distinct modes within a switching cycle, as depicted in Fig. 3, with the corresponding key waveforms shown in Fig. 4. The detailed mode-by-mode operation is explained in the following subsections.

A. *Mode 1 ($t_0 < t < t_1$):*

During this mode, the switch (Q) is turned on when an external pulse is applied to its gate terminal, as shown in Fig. 3. Inductor $L_1$ stores energy from the source voltage through diode $D_2$. Capacitors $C_4$ and $C_6$ discharge, which forward-biases diode $D_5$, allowing it to carry current and magnetize inductor $L_3$. At the end of this mode, the current through diode $D_5$ begins to decrease. The associated circuit and current waveforms for the inductors and diodes are shown in Fig. 3a and Fig. 4.

By applying Kirchhoff's Voltage Law (KVL), the following voltage equations can be derived:

$$V_{L_1} = V_{in} \; ; \; V_{L_2} = V_{C_1} \quad (1a)$$
$$V_{L_3} = -V_{C_2} + V_{C_3} = V_{C_4} - V_{C_6}; \quad (1b)$$
$$V_{C_3} + V_{C_5} + V_{C_6} = V_o \quad (1c)$$

It is worth noting that by applying Kirchhoff's Current Law (KCL) to the circuit nodes in mode 1, the following currents can be obtained:

$$i_{L_1} = i_{in}; \; i_{C_1} = -i_{L_2}; \; i_{C_2} = i_{L_3} + i_{D_5} \quad (2a)$$
$$i_{C_3} = -i_{D_5} - i_o - i_{L_3}; \; i_{C_4} = i_{D_5} \quad (2b)$$
$$i_{C_5} = -i_o; \; i_{C_6} = -i_{D_5} - i_o \quad (2c)$$

B. *Mode 2 ($t_1 < t < t_2$):*

In this mode, the switch Q remains on, and diode $D_2$ conducts to magnetize inductor $L_1$, while all other diodes remain off. This operation is shown in Fig. 3b. To obtain the inductor voltages, Kirchhoff's Voltage Law (KVL) can be applied to the circuit in Fig. 3b.

$$V_{L_1} = V_{in} \; ; \; V_{L_2} = V_{C_1} \quad (3a)$$
$$V_{L_3} = -V_{C_2} + V_{C_3}; \; V_{C_3} + V_{C_5} + V_{C_6} = V_o \quad (3b)$$

By applying Kirchhoff's Current Law (KCL), the capacitor currents can be obtained in the same manner as in mode 1, except for capacitor $C_4$, which is not involved in this mode.

$$i_{C_6} = -i_o \quad (4a)$$

C. *Mode 3 ($t_2 < t < t_3$):*

In mode 3, the switch is turned off, which forward-biases diode $D_1$, allowing current to flow from inductor $L_1$ to capacitor $C_1$. Capacitor $C_2$ discharges, causing diode $D_4$ to conduct. Meanwhile, capacitor $C_5$ is charged by $C_4$ through diode $D_6$. The current through diode $D_6$ decreases and eventually reaches zero at the end of this mode. This operation is depicted in Fig. 3c.

To determine the voltages in this mode, Kirchhoff's Voltage Law (KVL) is applied to the circuit in mode 3, resulting in the following voltage equations.

$$V_{L_1} = V_{in} - V_{C_1}; \; V_{L_2} = V_{C_1} + V_{C_2} - V_{C_3} - V_{C_6}; \quad (5a)$$
$$V_{L_3} = -V_{C_6}; V_{C_4} = V_{C_5} \quad (5b)$$
$$V_{C_3} + V_{C_5} + V_{C_6} = V_o \quad (5c)$$

Applying KCL in Fig. 3c, the following equations of currents are obtained:

$$i_{L_1} = i_{in} = i_{D_1}; \; i_{C_1} = i_{L_1} - i_{L_2}; \; i_{C_2} = -i_{L_2} \quad (6a)$$
$$i_{C_3} = i_{D_4} + i_{D_6} - i_{L_3} - i_o; i_{C_4} = -i_{D_6}; \quad (6b)$$
$$i_{C_5} = i_{D_6} - i_o; \; i_{C_6} = i_{D_4} + i_{D_6} - i_o \quad (6c)$$

D. *Mode 4 ($t_3 < t < t_4$):*

As shown in Fig. 3d, the switch remains off during this mode, and diodes $D_1$ and $D_3$ are forward-biased. Capacitor $C_3$ is charged by inductor $L_2$.

By applying KVL, the following voltages are obtained.

$$V_{L_1} = V_{in} - V_{C_1}; \; V_{L_2} = V_{C_1} - V_{C_3} \quad (7a)$$
$$V_{L_3} = -V_{C_2}; \; V_{C_3} + V_{C_5} + V_{C_6} = V_o \quad (7b)$$

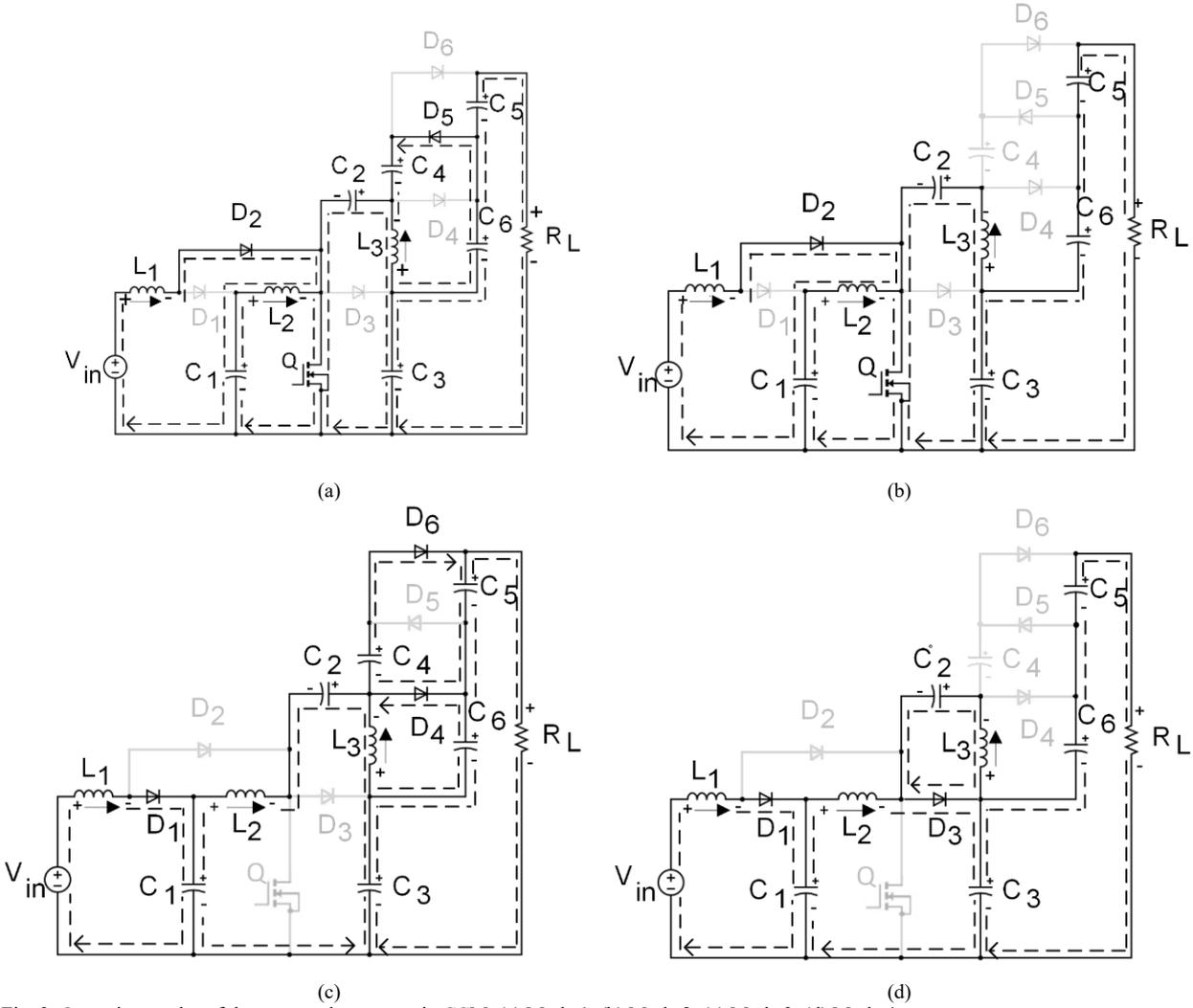

(a)　　　　　　　　　　　　　　　　(b)

(c)　　　　　　　　　　　　　　　　(d)

Fig. 3. Operation modes of the proposed converter in CCM. (a) Mode 1. (b) Mode 2. (c) Mode 3. (d) Mode 4.

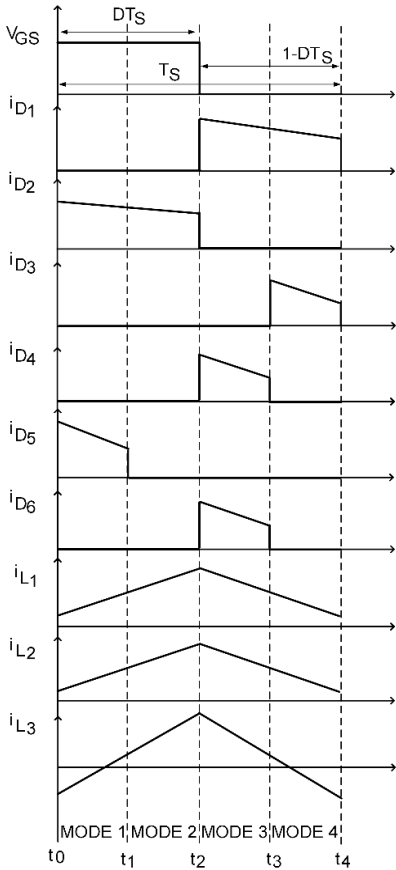

Fig. 4. Waveforms of diode and inductor currents during different modes of operation of proposed modified QBC in CCM mode.

The current equations can be obtained by applying KCL in the circuit in Fig. 3d.

$$i_{L_1} = i_{in} = i_{D_1};\ i_{C_1} = i_{L_1} - i_{L_2} \qquad (8a)$$
$$i_{C_2} = i_{L_3};\ i_{C_3} = i_{L_2} - i_o; \qquad (8b)$$
$$i_{C_5} = i_{C_6} = -i_o;\ i_{D_3} = i_{L_2} + i_{L_3} \qquad (8c)$$

The capacitor voltages can be derived using the volt-second balance principle on each inductor, considering one off-state mode and one on-state mode. Mode 2 is ignored due to its short duration.

By applying the volt-second balance principle to inductor $L_1$ in equations (1a) and (5a), the voltage of capacitor $C_2$ can be computed as follows:

$$\langle V_{L_1} \rangle = V_{in}DT_s + (V_{in} - V_{C_1})(1-D)T_s = 0$$
$$V_{in}D + (V_{in} - V_{C_1})(1-D) = 0$$
$$V_{C_1}(1-D) = V_{in}$$
$$V_{C_1} = \frac{V_{in}}{1-D} \qquad (9)$$

Similarly, applying the volt-second balance on inductors $L_2$, $L_3$ and $L_4$, the below equations can be obtained.

$$V_{C_2} = V_{C_6} = \frac{V_{in}D}{(1-D)^2} \qquad (10a)$$
$$V_{C_3} = V_{C_4} = V_{C_5} = \frac{V_{in}}{(1-D)^2} \qquad (10b)$$

where $D$ is the duty cycle and $T_s$ is the time-period.

The gain of the proposed converter during CCM is derived by computing the ratio of output voltage to input voltage. The gain of the proposed converter is higher than that of the classical QBC converter which was derived to be $\frac{1}{(1-D)^2}$.

$$M_{CCM} = \frac{V_o}{V_{in}} = \frac{V_{C_3}+V_{C_5}+V_{C_6}}{V_{in}} = \frac{\frac{V_{in}}{(1-D)^2}+\frac{V_{in}}{(1-D)^2}+\frac{V_{in}D}{(1-D)^2}}{V_{in}}$$

$$M_{CCM} = \frac{2+D}{(1-D)^2} \quad (11)$$

The voltage stress across the switch is derived when the switch is off. Hence, considering mode 3, the voltage stress across the switch is found as

$$V_Q = V_{C_1} = \frac{V_{in}}{(1-D)^2} = \frac{V_o}{2+D} \quad (12)$$

The voltage stress across the diodes is derived to be:

$$V_{D_1} = -\frac{V_{in}}{1-D} = -\frac{V_o(1-D)}{2+D} \quad (13a)$$

$$V_{D_2} = -\frac{V_{in}D}{(1-D)^2} = -\frac{V_oD}{2+D} \quad (13b)$$

$$V_{D_3} = V_{D_4} = V_{D_5} = V_{D_6} = -\frac{V_{in}}{(1-D)^2} = -\frac{V_o}{2+D} \quad (13c)$$

It is evident from equations (12) and (13) that the voltage and diode stresses are significantly reduced compared to the classical QBC. In the classical QBC, the switch stress is equal to $\frac{V_o}{1-D}$, while the diode stresses are equal to $V_o$ and $\frac{V_o}{1-D}$.

## III. CONVERTER DESIGN AND COMPONENT SELECTION

This section presents the design of the key components for the proposed converter, including inductors and capacitors, to ensure stable operation in continuous conduction mode (CCM).

### A. Inductor, Diode, and Switch Currents

To determine the values of the inductors and capacitors, the average currents through these components are first computed. The average current through inductor $L_1$ can be expressed using equations (2a) and (11), which correspond to the average input current:

$$I_{L_1} = I_{in} = I_o M_{CCM} \quad (14)$$

The inductor currents are obtained using the ampere-second balance principle, which states that the average current through a capacitor in steady state is zero. Applying this principle to capacitor $C_1$ in equations (2a) and (6a), the average current through inductor $L_2$ is derived as:

$$\langle I_{C_1} \rangle = -I_{L_2}DT_s + (I_{L_1}-I_{L_2})(1-D)T_s = 0$$
$$I_{L_2} = I_{L_1}(1-D)$$
$$I_{L_2} = I_o M_{CCM}(1-D) \quad (15)$$

The average currents of the remaining inductors and diodes can be derived in a similar manner using the same principles, following the steady-state and ampere-second balance conditions which are as follows:

$$I_{L_3} = \frac{I_o(3+D)}{D} \; ; \; I_{D_1} = I_o M_{CCM}(1-D) \quad (16a)$$

$$I_{D_2} = I_o M_{CCM}D; \; I_{D_3} = 3I_o \quad (16b)$$

$$I_{D_4} = I_{D_5} = I_{D_6} = I_o; \; I_Q = I_o M_{CCM}(2-D) \quad (16c)$$

### B. Design of Inductors

To maintain CCM, the peak-to-peak inductor ripple current is assumed to be twice the average inductor current:

$$\Delta i_{L_1} = 2I_{L_1} \quad (17)$$

where $\Delta i_{L_1}$ is the peak-to-peak ripple current of inductor $L_1$. The ripple current can also be expressed using the standard inductance formula:

$$\Delta i_{L_1} = \frac{V_{in}D}{L_1 f_s} \quad (18)$$

where $f_s$ is the switching frequency. On substituting (14) and (17) into (18), the required inductance for $L_1$ is obtained as:

$$L_1 = \frac{V_{in}D}{2I_o M_{CCM}f_s} \quad (19)$$

To guarantee CCM operation, the minimum inductance is:

$$L_1 \geq \frac{V_{in}^2 D}{2 P_o f_s} \quad (20)$$

where $P_o = V_o I_o$ is the converter output power. Similarly, the minimum inductances of the remaining inductors are:

$$L_2 \geq \frac{V_{in}^2 D}{2(1-D)^2 P_o f_s}; \; L_3 \geq \frac{V_{in}^2 D^2 M_{CCM}}{2(1-D)(3+D)P_o f_s} \quad (21)$$

### C. Design of Capacitors

The capacitances are computed based on the average current through each capacitor. For capacitor $C_1$, the current is derived from (2) as:

$$I_{C_1} = -I_o M_{CCM}(1-D) \quad (22)$$

Using the standard capacitor formula, the minimum value of $C_1$ is:

$$C_1 \geq \frac{I_{C_1}D}{\Delta v_{C_1} f_s}$$
$$C_1 \geq \frac{I_o M_{CCM} D(1-D)}{\Delta v_{C_1} f_s} \quad (23)$$

where $\Delta v_{C_1}$ is typically chosen as 10% of the nominal capacitor voltage. Ssimilarly, the minimum capacitances for the remaining capacitors are:

$$C_2 \geq \frac{I_o M_{CCM}(1-D)^2}{\Delta v_{C_2} f_s}; \; C_3 \geq \frac{I_o D}{\Delta v_{C_3} f_s} \quad (24a)$$

$$C_4 \geq \frac{I_o D}{\Delta v_{C_4} f_s}; \; C_5 \geq \frac{I_o D}{\Delta v_{C_5} f_s}; \; C_6 \geq \frac{I_o(1-D)}{\Delta v_{C_6} f_s} \quad (24b)$$

These calculated inductances and capacitances ensure CCM operation and maintain low voltage/current stress across the semiconductor devices.

## IV. COMPARATIVE PERFORMANCE VALIDATION

This section highlights the advantages of the proposed converter over existing step-up topologies and validates its performance improvements through MATLAB/Simulink simulations and a hardware prototype, supported by recent literature. The results demonstrated the superior performance of the proposed converter are discussed in detail below.

### A. Comparison with Existing Converter Topologies

The proposed HGVM-QBC is compared with converters reported in [27], [28], [29], [31], [32], as well as the conventional quadratic boost converter (QBC), as illustrated in Fig. 5. The comparison criteria include voltage gain, switch voltage stress, and diode voltage stress.

From Fig. 5(a), it is evident that the voltage gain of the proposed converter surpasses that of the compared topologies, particularly for duty ratios above 0.4. Although a few converters achieve higher gain for D < 0.4, the absolute voltage gain remains low and insufficient for high-voltage applications.

The converter reported in [31] exhibits higher gain for D < 0.33, but it employs two switches, resulting in higher switch voltage stress. Fig. 5(b) shows the switch voltage stresses of HGVM-QBC and other converters. The proposed HGVM-QBC demonstrates the lowest switch stress as the duty cycle increases. While the converters in [27] and [28] also achieve relatively low switch stress, they are inferior in terms of voltage gain and require additional switches, which contribute to higher stress. The converter in [32] maintains the lowest constant switch stress among the considered topologies, but its voltage gain is limited.

Fig. 5c presents the diode voltage stresses, showing that the HGVM-QBC experiences lower diode stress compared to other converters. The converter in [27] exhibits significantly higher diode stresses and is excluded from the plot for better visibility of the remaining data.

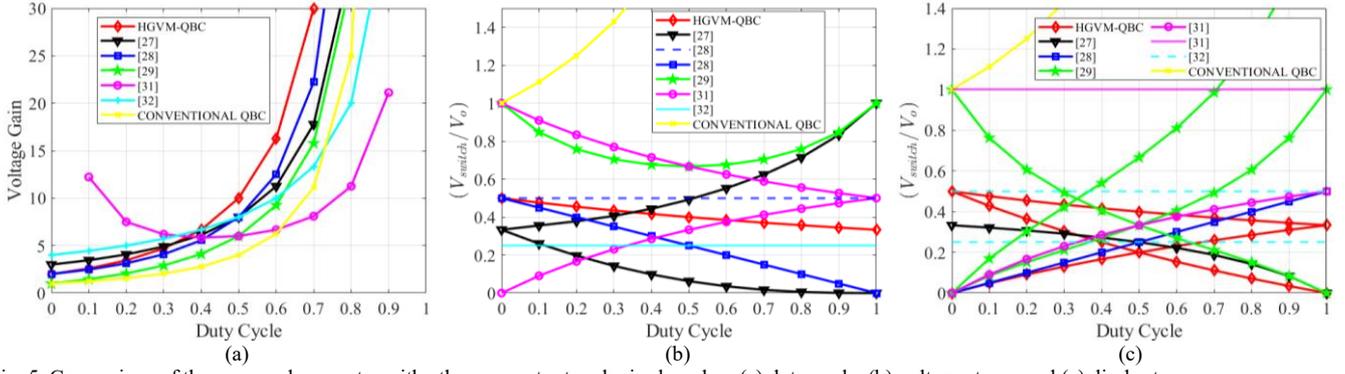

Fig. 5. Comparison of the proposed converter with other converter topologies based on (a) duty cycle, (b) voltage stress, and (c) diode stress.

TABLE I
COMPARISON OF PROPOSED HGVM-QBC WITH OTHER CONVERTER TOPOLOGIES

| Topology | Switch | Diode | Capacitor | Inductor | Voltage Gain | Switch Voltage Stress | Diode Voltage Stress |
|---|---|---|---|---|---|---|---|
| HGVM-QBC | 1 | 6 | 6 | 3 | $\dfrac{2+D}{(1-D)^2}$ | $\dfrac{V_o}{2+D}$ | $\dfrac{V_o(1-D)}{2+D}, \dfrac{V_o D}{2+D}, \dfrac{V_o}{2+D}$ |
| Conventional QBC | 1 | 3 | 2 | 2 | $\dfrac{1}{(1-D)^2}$ | $\dfrac{V_o}{1-D}$ | $\dfrac{V_o}{1-D}, V_o$ |
| [27] | 2 | 3 | 3 | 2 | $\dfrac{3-2D}{(1-D)^2}$ | $\dfrac{V_o(1-D)^3}{3-2D},$ $V_o - \dfrac{2V_o D(1-D)^2}{3-2D}$ | $\dfrac{V_o(1-D)}{3-2D}, \dfrac{V_o(3D-8)}{3-2D},$ $\dfrac{V_o(-6D^3+19D^2-18D+9)}{2D^3-5D^2+3D}$ |
| [28] | 2 | 4 | 3 | 2 | $\dfrac{2}{(1-D)^2}$ | $\dfrac{V_o}{2}, \dfrac{V_o(1-D)}{2}$ | $\dfrac{V_o D}{2}$ |
| [29] | 1 | 5 | 6 | 4 | $\dfrac{1+2D-2D^2}{(1-D)^2}$ | $\dfrac{V_o}{1+2D-2D^2}$ | $\dfrac{V_o(1-D)}{1+2D-2D^2}, \dfrac{V_o D}{1+2D-2D^2},$ $\dfrac{V_o D^2}{1+2D-2D^2}$ |
| [31] | 2 | 3 | 3 | 2 | $\dfrac{1+D}{D(1-D)}$ | $\dfrac{V_o D}{1+D}, \dfrac{V_o}{1+D}$ | $\dfrac{V_o D}{1+D}, V_o$ |
| [32] | 2 | 4 | 4 | 2 | $\dfrac{4}{1-D}$ | $\dfrac{V_o}{4}$ | $\dfrac{V_o}{2}, \dfrac{V_o}{4}$ |

Table I presents a detailed comparison of the proposed HGVM-QBC with existing converter topologies, considering key parameters such as the number of capacitors, inductors, diodes, and switches, as well as voltage gain, switch voltage stress, and diode stress. Notably, the proposed converter demonstrates significant improvements over the classical QBC and other recent designs, offering higher voltage gain while simultaneously reducing voltage stress on switches and diodes.

Overall, the comparison confirms that the HGVM-QBC provides a well-balanced enhancement in both voltage gain and semiconductor stress reduction, making it highly suitable for small-scale photovoltaic applications.

### B. Performance Comparison of the Proposed Converter and Classical QBC

This subsection examines and compares the operational performance of the proposed HGVM-QBC with a recent converter reported in [27], focusing on voltage gain and switch/diode voltage stresses. The enhanced characteristics of the proposed converter are validated using MATLAB/Simulink simulations and an experimental prototype.

An input voltage of $V_{in} = 12V$ is stepped up to an output voltage of $V_o = 151V$, with a load resistance of $R = 500\Omega$ and an output power $P_o = 200W$. The switching frequency is set to $f_s = 50kHz$. Using (11), the voltage gain and duty ratio are computed as $M_{CCM} = 12.59$ and $D = 0.55$. giving the output current $I_o = P_o/V_o = 1.32A$.

Substituting these values into (20) and (21), the critical inductance values for CCM operation are:
$$L_1 \geq 3.96\mu H \;;\; L_2 \geq 19.55\mu H;\; L_3 \geq 17.16\mu H \quad (25)$$
For improved step-up performance, the chosen inductance values are well above the critical values: $L_1 = 250\mu H, L_2 = 90\mu H$, and $L_3 = 82\mu H$.

The capacitor values are computed using (23) and (24), assuming a voltage ripple of 10%:
$$\Delta v_{C_1} = 0.1 \times \dfrac{V_{in}}{1-D} = 0.1 \times \dfrac{12}{1-0.5} = 2.6V \quad (26a)$$
$$C_1 \geq \dfrac{I_o M_{CCM} D(1-D)}{\Delta v_{C_1} f_s} \geq 31.6\mu F \quad (26b)$$

Similarly, the remaining capacitances are determined as:
$$C_2 \geq 22\mu F \;;\; C_3 \geq 16.5\mu F \quad (27a)$$
$$C_4, C_5 \geq 2.5\mu F; C_6 \geq 3.8\mu F \quad (27b)$$

For experimental setup, the chosen capacitors are: $C_1 = 50\mu F, C_2, C_4, C_5, C_6 = 80\mu F, C_3 = 110\mu F$.

For comparison, the converter in [27] achieves a voltage gain of $M_{CCM} = 9.382$, yielding $V_o = 112.6V$. Its inductors are chosen close to the critical values: $L_1 \geq 425\mu H \;;\; L_2 \geq 26.1\mu H$; selected as $L_1 = 425\mu H \;;\; L_2 = 27\mu H$. The corresponding capacitors are: $C_1 = 4\mu F, C_2 = 6\mu F$ and $C_3 = 100\mu F$, assuming a 10% voltage ripple.

Simulation results are shown in Fig. 6. As illustrated in Fig. 6(a), the HGVM-QBC steps up the input voltage to 151 V, whereas the converter in [27] produces 112.5 V under the same conditions. Fig. 6(b) compares the switch voltage stresses, showing that the proposed converter reduces the MOSFET stress to approximately 60 V, closely matching theoretical predictions from (12).

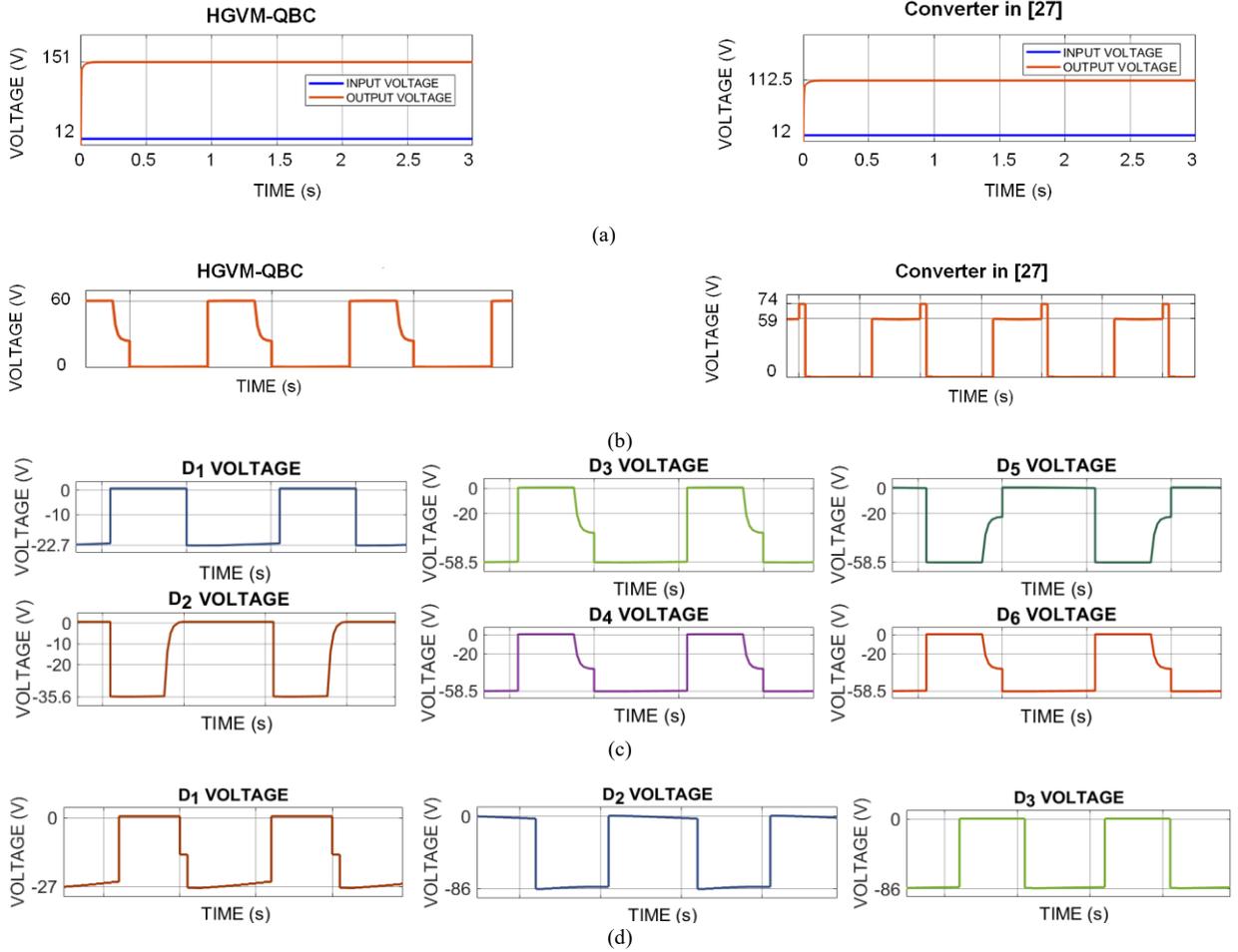

Fig. 6. MATLAB/Simulink simulation results of the proposed HGVM-QBC and the converter in [27] based on (a) input & output voltages (b) voltage stress on switch and (c) diode voltage stresses of HGVM-QBC, (d) Diode voltage stresses from the converter in [27].

Figs. 6(c) and 6(d) illustrate the diode voltage stresses. The HGVM-QBC exhibits significantly lower stresses: $V_{D1}\approx22.7$ V, $V_{D2}\approx35.6$ V, $V_{D3-D6}\approx58.5$ V, which aligns closely with theoretical values from (13). In contrast, the converter in [27] experiences higher diode stresses of 27 V and 86 V.

These results confirm that the proposed HGVM-QBC achieves higher voltage gain and lower semiconductor stresses compared to recent topologies, validating its suitability for small-scale PV applications.

### C. Performance Validation via Experimental Setup

To demonstrate the performance and feasibility of the proposed HGVM-QBC, a 200 W prototype was developed as shown in Fig. 7 and tested in the laboratory with closed-loop voltage control shown in Fig. 8. To mitigate the reverse recovery effect in the diodes, Schottky diodes (C3D20060) were employed in the design. The detailed component specifications of the prototype are listed in Table II.

Fig. 9 shows the voltage tracking performance of the proposed converter with closed-loop PI controller via duty ratio D. The output dc-voltage is measured and fed back to the controller to directly generate the duty ratio for the single switch Q. The control voltage performance of tracking difference step-change scenarios is shown, including low-gain region with voltage tracking from 40[V] to 60[V] in Fig. 9(a). The precise voltage regulation at steady-state error 0.2% is measured, with small ripples smaller than 0.1 V, and the relevant duty-ratio at 0.2 and 0.3, respectively. The smooth transition with no overshoot and settling time of voltage at 3 [ms] only. At high-gain regions shown in Figs. 9 (b) & (c) when output voltage reaches 150V, higher ripples are observed smaller than 0.65V around the steady-state values with respect to the duty-cycle of 0.55 as seen in Fig. 9(c). Afterall, the dc/dc converter responses well with the voltage regulation by a simple PI control structure and showing its high gain performance with stable voltage and input duty-ratio as expected from the theory analysis.

Experimental results are illustrated in Fig. 10, showing the input-output voltages, as well as the voltages and currents of the capacitors, diodes, switches, and inductors. Fig. 10(a) confirms that the 12 V input is successfully stepped up to 151 V output. The switch and diode voltage stresses are depicted in Figs. 10(b) and 10(c), respectively. The measured MOSFET voltage stress is approximately 61 V, which is in close agreement with the theoretical prediction from (12) and the MATLAB/Simulink simulation results. These results validate the high-gain capability and low-voltage stress performance of the proposed converter, confirming its suitability for small-scale photovoltaic applications.

The voltage stresses across diodes were measured as approximately: $V_{D1}\approx-25.6$ V, $V_{D2}\approx-40$ V, $V_{D3-D6}\approx-60$ V. These values closely match the theoretical predictions from (13) and MATLAB simulation results.

Capacitor voltages are illustrated in Figs. 10(d) and 10(e). The measured values are: $V_{C1}=23.7$ V, $V_{C2}=33.8$ V, $V_{C3}=57$ V, $V_{C4}=57.6$ V, $V_{C5}=56.8$ V, $V_{C6}=35.5$ V. Notably, $V_{C2}\approx V_{C6}$ and $V_{C3}\approx V_{C4}\approx V_{C5}$, consistent with theoretical values from (9) and (10). Minor deviations arise due to parasitic capacitances and inductances, which dampen voltage spikes.

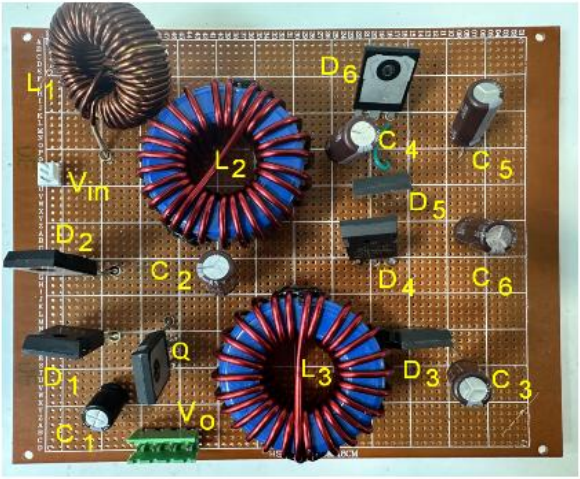

Fig. 7. Hardware prototype of proposed HGVM-QBC.

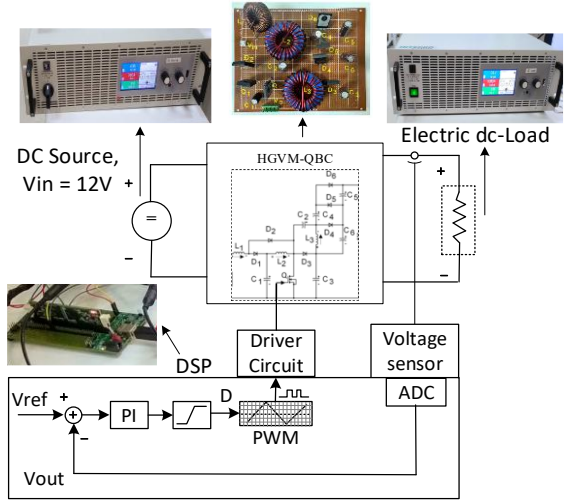

Fig. 8. Closed loop voltage control circuit of proposed HGVM-QBC.

TABLE II
HARDWARE SPECIFICATIONS OF THE PROPOSED HIGH GAIN & LOW STRESS CONVERTER

| Components | Specification |
| --- | --- |
| Input Voltage ($V_{in}$) | 12V |
| Output Voltage ($V_o$) | 151V |
| Output Power | 200W |
| Load ($R_L$) | 500Ω |
| Switching Frequency ($f_s$) | 50kHz |
| Duty Cycle ($D$) | 55% |
| Capacitors | $C_1 = 110\mu F, C_2 - C_6 = 220\mu F$ |
| Inductors | $L_1 = 250\mu H, L_2 = 90\mu H, L_3 = 80\mu H$ |
| Switch | UJ3C120040K3S |
| Diodes | $D_1$: 30CPH03, $D_2 - D_6$: C3D20060 |

Inductor currents are shown in Fig. 10(f), with measured ripple values: $\Delta I_{L1}$=0.412 A, $\Delta I_{L2}$=2.7 A, $\Delta I_{L3}$=2.95 A. These ripples are lower than the maximum theoretical values derived from (14)–(17), due to parasitic effects and switching losses.

Switch and diode currents are depicted in Figs. 10(g) & (h): $I_{Q1}$=12 A, $I_{D1}$=5.5 A, $I_{D2}$=5 A, $I_{D3, D5}$=3 A, $I_{D4}$=3.5 A, $I_{D6}$=5 A

These values are slightly lower than the theoretical predictions from (16) due to parasitic effects, while Schottky diodes reduce the impact of reverse recovery currents.

Overall, the experimental results confirm the effectiveness of the HGVM-QBC, demonstrating high voltage step-up (12 V to 151 V), low semiconductor voltage stress, and efficient CCM operation, validating the theoretical and simulation analyses.

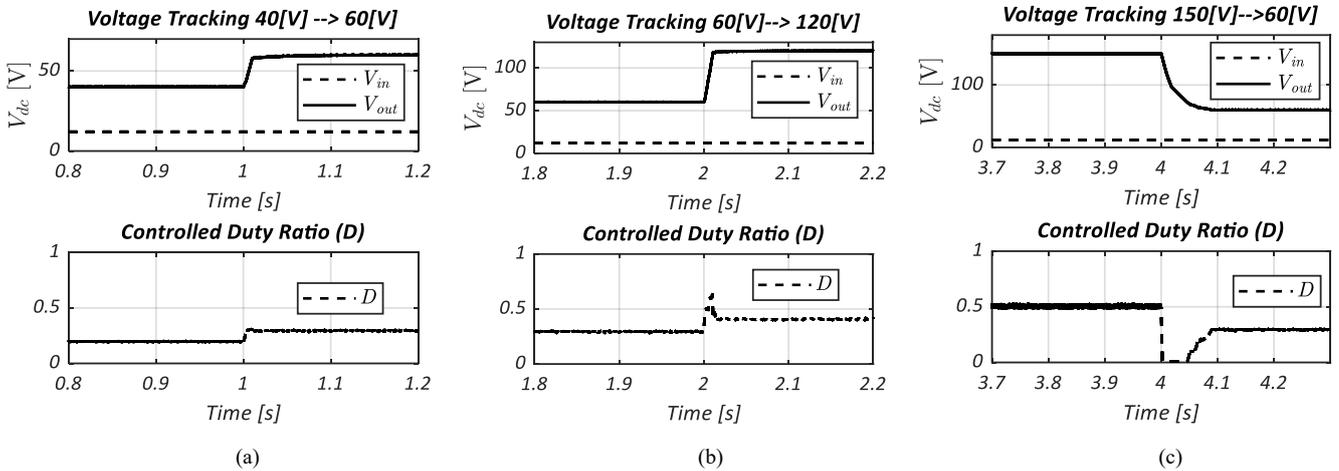

Fig. 9. Closed-loop output-voltage tracking with step-changed references of the proposed HGVM-QBC with conventional PI control design, (a) low-gain range of [40- 60] V output, (b) high-gain range of [60 - 120] V, (c) high-gain, large-step change from 150 [V] to 60 [V].

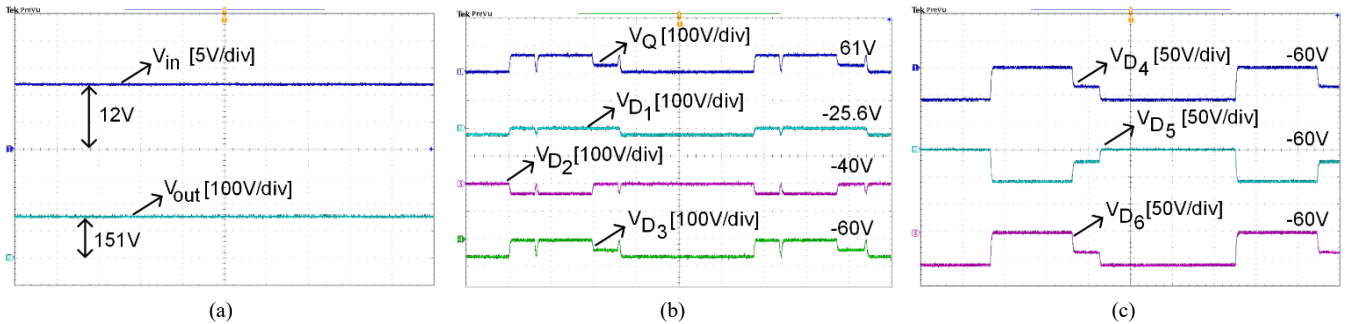

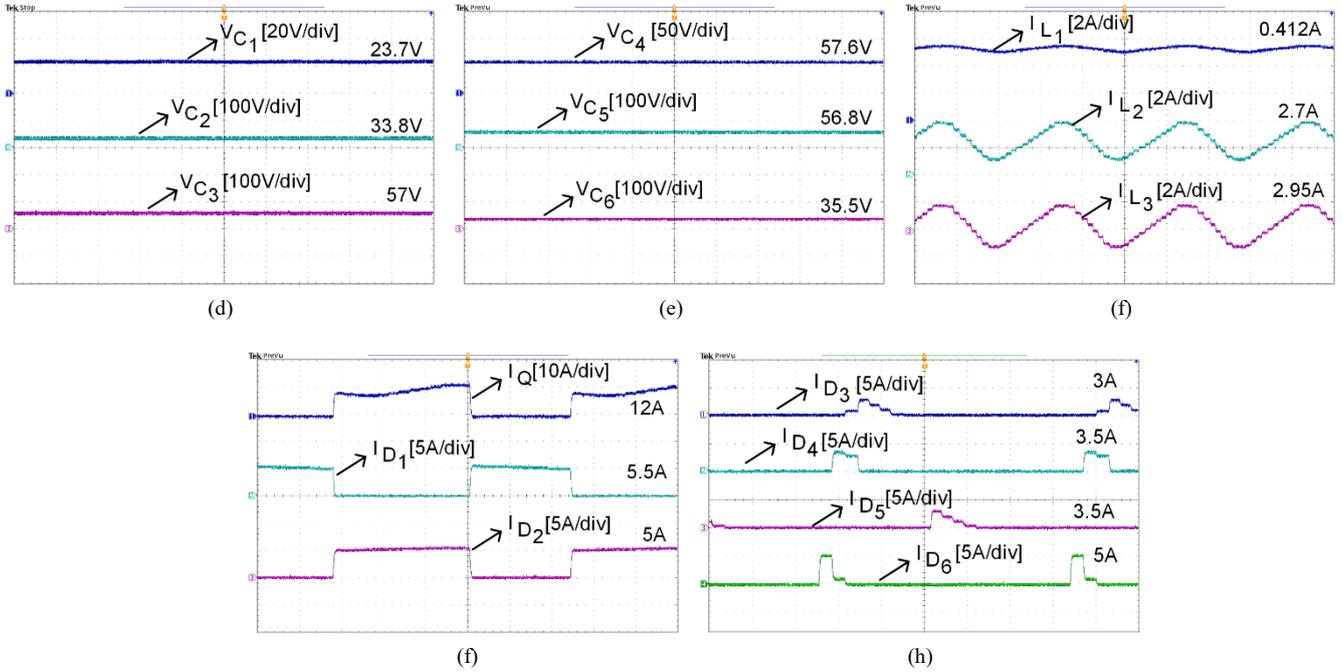

Fig. 10. Hardware results of the proposed HGVM-QBC at 55% duty cycle. (a) Input and output voltages measured at 4µs/div. (b) Voltages of switch and diodes $D_1$, $D_2$ and $D_3$, at 20µs/div. (c) Voltages of diodes $D_4$, $D_5$, and $D_6$ at 20µs. (d) Voltages of capacitors $C_1$, $C_2$ and $C_3$ at 40µs/div. (e) Voltages of capacitors $C_4$, $C_5$ and $C_6$ at 40µs/div. (f) Inductor currents at 40µs/div. (g) Currents of switch and diodes $D_1$, $D_2$ and $D_3$, at 20µs/div. (h) Currents of diodes $D_4$, $D_5$, and $D_6$ at 20µs/div.

## V. CONCLUSION

This paper proposes a single-switch high-gain voltage-multiplier coupled quadratic boost converter (HGVM-QBC) derived from the conventional QBC. The proposed topology addresses key challenges in DC-DC conversion by providing high voltage gain, reduced semiconductor voltage stress, and continuous conduction mode (CCM) operation. By incorporating a voltage multiplier cell into the classical QBC, the converter achieves enhanced voltage gain while maintaining low stress on both switches and diodes, improving overall efficiency and reliability. Comparative analysis with conventional QBCs and recent converters demonstrates that the HGVM-QBC exhibits superior performance in terms of voltage gain, switch voltage stress, and diode stress, especially at higher duty cycles. The key contributions include the development of a compact, efficient converter topology that operates at low duty cycles, reduces component stresses, and enhances voltage gain, making it highly suitable for small-scale PV systems and other applications with space and cost constraints. Future work may focus on further optimization of the converter's performance and its integration into practical renewable energy systems, including custom designs for electric vehicles and distributed energy applications.